\documentclass[prb,twocolumn,amssymb,showpacs,floatfix]{revtex4}
\usepackage{amsmath}
\usepackage{graphicx}

\begin{document}
\title{Direct evidence of soft mode behavior near the Burns' temperature in
PbMg$_{1 / 3}$Nb$_{2 / 3}$O$_{3}$ (PMN) relaxor ferroectric}
\author{S. B. Vakhrushev}
\affiliation{ A. F. Ioffe Physico-Technical Institute, St.-Petersburg 194021, Russia.}%}
\email{s.vakhrushev@pop.ioffe.rssi.ru}
\author{S. M. Shapiro}
\affiliation{Physics Department, Brookhaven National Laboratory,
Upton, NY 11973}
\date{today}

\begin{abstract}

Inelastic neutron scattering measurements of the relaxor
ferroelectric PbMg$_{1 / 3}$Nb$_{2 / 3}$O$_{3}$ (PMN) in the
temperature range 490~K$<$T$<$880~K directly observe the soft mode
(SM) associated with the Curie-Weiss behavior of the dielectric
constant $\varepsilon $(T). The results are treated within the
framework of the coupled SM and transverse optic (TO1) mode and
the temperature dependence of the SM frequency at q=0.075 a* is
determined. The parameters of the SM are consistent with the
earlier estimates and the frequency exhibits a minimum near the
Burns temperature ($\approx $ 650K)
\end{abstract}


\maketitle

\bigskip

\section{Introduction}\label{sec:intro}

One of the most interesting group of disordered compounds
undergoing structural phase transitions is the so-called relaxor
ferroelectrics \cite{1,2}, which have ferroelectric-like
properties with diffuse phase transitions. Relaxor ferroelectrics
(RF) constitute a large group of disordered perovskite-like
crystals. The main feature of RF `s is related to the completely
or partially random occupation of the equivalent positions by
different (usually nonisovalent) ions. This chemical disorder
results in the destruction of the normal ferroelectric phase
transition and the appearance of new physical properties similar
to those of disordered magnets such as spin glasses. There are now
hundreds of RF's and most of them have the simple cubic perovskite
structure

The lead magnoniobate PbMg$_{1 / 3}$Nb$_{2 / 3}$O$_{3}$ (PMN) is
considered a model system for the cubic RF's. It has been
extensively studied and it was shown that PMN exhibits a
glass-like phase transition to a non-ergodic phase at a freezing
temperature T$_{g}\approx$230K as determined from the appearance
of history dependent effects and the divergence of the nonlinear
susceptibility \cite{3,4}. This transition is accompanied by the
appearance of the broad frequency dependent peak in the
temperature dependence of the dielectric constant, $\varepsilon
(T)$. No apparent symmetry changes were detected either by X-ray
or neutron scattering experiments \cite{5} and no long range
ferroelctric order appears at low temperature \cite{6}. It was
also demonstrated that besides T$_{g}$ another singular
temperature point exists \cite{7}, the so-called Burns
temperature, T$_{d}$, which in PMN is $\approx $650~K. Above
T$_{d}$ PMN and other relaxors behave like normal ferroelctrics,
while below this temperature there are clear indications of the
local lowering of the symmetry and formation of some polar
nanoregions. In case of Pb(Zr$_{1 - x}$Ti$_{x})$O$_{3}$ (PZT) and
related compounds, T$_{d}$ can be attributed to a real structural
phase transition where the crystal transforms from a cubic to a
lower symmetry. However, in the case of PMN, the situation is less
clear because there is no reduction in symmetry. It is important
to emphasize, that above T$_{d}$, $\varepsilon (T)$ follows
Curie-Weiss law \cite{33} with T$_{0}$=398~K. In Ref.
\onlinecite{18} an analysis of the ultrabroadband dielectric
spectroscopy data is presented and demonstrated that the phonon
contribution to the permittivity is dominant and determines the
Curie-Weiss behavior above T$_{d}$.

There exists limited information about the lattice dynamics of
RF's. Several Raman and infrared studies \cite{8,9,10,11,13,14} of
the q=0 modes of PMN and other relaxors were reported, but no
adequate understanding of the lattice dynamics has been achieved.
A detailed neutron inelastic scattering experiment on PMN was
undertaken \cite{15}. For temperatures above T$_{d}\approx$650~K,
there is a broad inelastic response whose intensity increases as
the temperature decreases. This scattering was attributed to the
defect induced optic-like mode, called the quasi-optic (QO) mode,
which couples strongly with the transverse acoustic (TA) mode. It
was demonstrated that at T$_{d} \approx$650~K a crossover in the
critical dynamics takes place, resulting in a cessation of the
growth of the inelastic scattering intensity and the appearance of
a narrow (resolution limited) central peak. From the analysis of
the lineshapes of the phonon resonance related to the coupled
TA-QO modes, the estimation of the parameters of the QO mode was
made. The authors also predicted that the QO mode should be
directly observable in the narrow q-region near the Brillouin zone
(BZ) center. The results of the soft mode search in PMN were
reported \cite{16} at room temperature where it was demonstrated
that at a small q there is an increased damping of a transverse
optic mode, TO1, which has the appearance of a ``waterfall'' in
the dispersion curve. More recently \cite{17} a weak temperature
dependence of the TO1 was observed which varied as $\omega
^{2}\sim $(T-T$_{0})$ with negative T$_{0} \approx $-330K. This
mode was earlier assigned \cite{15}as the hard TO branch, but its
temperature dependence cannot explain the Curie-Weiss law above
T$_{d}$ so is not related to the high-temperature dielectric
properties of PMN. The aim of the present paper is to observe
directly a soft mode behavior associated with the Curie-Weiss law
in a PMN single crystal. We shall demonstrate that the QO mode
deduced in the earlier experiment is indeed the soft mode (SM)
associated with the high-temperature $\varepsilon $(T) dependence.

\section{Experiment}\label{sec:exper}

All the scattering experiments were performed with the same single
crystal used in Ref.~\onlinecite{15}. The crystal was grown by the
Czochralski technique in the Physical Institute of the University
of Rostov-on-Don. It was approximately triangular pyramid shape
with a volume of $\approx $0.5~cm$^{3}$. The crystal was of good
quality with a uniform mosaic spread of less than 40''. The
lattice parameter at room temperature is 4.04~{\AA}. The sample
was mounted in a vacuum furnace and the measurements performed at
temperatures 490~K $<$~T~$<$~880~K, i.e. above and below the Burns
temperature T$_{d}$.

The inelastic neutron measurements were performed on the H4M
three-axis spectrometer at Brookhaven's High Flux Beam Reactor.
The spectrometer was operated with a fixed final neutron
wavevector k$_{F}$= 2.662~{\AA}$^{ - 1}$ (E$_{F}$ = 14.7 meV).
Pyrolitic graphite crystals with a 25' mosaic spread were used
both as a monochromator and analyzer. The collimations were
40'-40'-60'-60', and the resulting energy resolution
(Full-width-at-Half Maximum) was $\approx$1.0~meV. The instrument
parameters were chosen to obtain a good compromise between
scattered intensity, resolution and the possibility to perform the
measurements in several Brillouin zones. The sample was mounted
with [1 0 0] axis vertical. Scans were performed in the
constant-\textbf{Q} mode, i.e. the scattering vector \textbf{Q}
was kept constant while the energy was varied.

\section{Experimental results}\label{sec:res}

\begin{figure}
\includegraphics [width=\columnwidth,clip=] {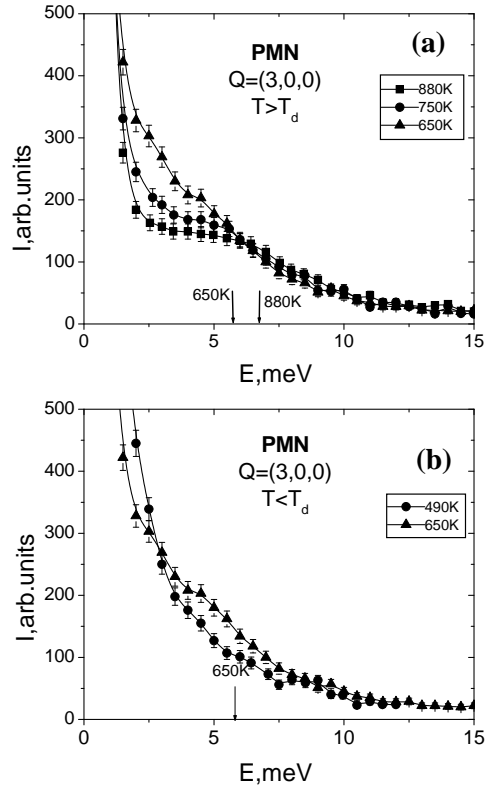}
\caption{Inelastic neutron scattering curves in PMN at Q=(3,0,0)
at temperatures above (a) and below (b) T$_{d}$, The lines
corresponds to fit by Eq. (1); arrows corresponds to the
frequencies of TO1 mode at 880K and 650K are from
Ref.~\protect{\onlinecite{17}}.} \label{fig:1}
\end{figure}

\begin{figure}
\includegraphics [width=\columnwidth,clip=] {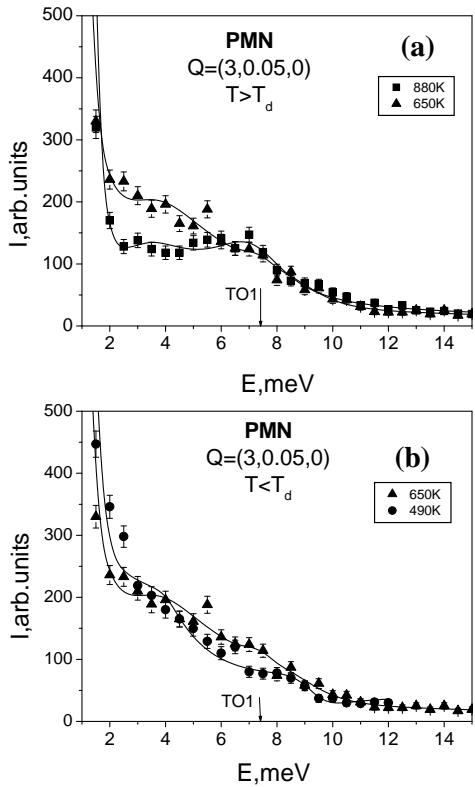}
\caption{Inelastic neutron scattering curves in PMN at
Q=(3,0.05,0) at temperatures above (a) and below (b) T$_{d}$, The
lines corresponds to fit by Eq. (1); arrows corresponds to the
frequency of TO1 mode from Ref.~\protect{\onlinecite{17}}.}
\label{fig:2}
\end{figure}

\begin{figure}
\includegraphics [width=\columnwidth, height= \columnwidth, clip=] {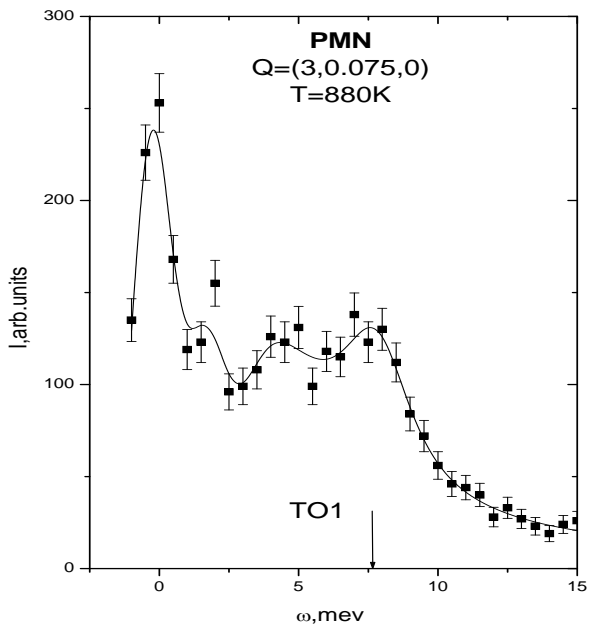}
\caption{Inelastic neutron scattering curves in PMN at
Q=(3,0.075,0) at 880K. The lines corresponds to fit by Eq~(1).}
\label{fig:3}
\end{figure}

We have used the previously obtained estimates of the soft mode
dispersion and damping to chose the position in reciprocal space
for observing the soft mode. Most of the measurements were
performed for \textbf{q} in the [010] direction near the (3,0,0)
reciprocal lattice vector. This contrasts with the recently
published results where measurements were mostly performed in the
Brillouin zone centered at (2,0,0). In the (3,0,0) Brillouin zone,
the intensity of the TA phonons is very low which would make
observation of the soft mode easier. The TO1 mode with frequency
at the zone center of about 7meV is also observable in this BZ.
Measurements were performed for q=0, 0.05, 0.075 and 0.1 in a*
units. Figures \ref{fig:1} and \ref{fig:2} show the inelastic
scans near (3,0,0) for temperatures above (a) and below (b)
T$_{d}$.
For temperatures above T$_{d}$=650K, cooling results in a strong
increase of the scattered intensity at the q=0 zone center for
small energy transfers (Fig.~\ref{fig:1}a). The arrows in
Fig.~\ref{fig:1}a, taken from Ref.~\onlinecite{17}, show the
positions of the TO1 mode, which changes very little in frequency
and intensity over the temperature range shown. The growth of
intensity at energies less that TO1 indicate that another mode is
present and its frequency is decreasing as the temperature
decreases. The spectra measured in this Brillouin Zone are
distinctly different than the observations of Ref. \onlinecite{17}
measured in the (2,0,0) zone. A similar, but less pronounced
effect is observed at q=0.05 a* as shown in Fig.~\ref{fig:2}a. For
larger q=0.075a* there is barely any temperature dependence in the
low energy region (Fig.~\ref{fig:3}).
On cooling below 650K the effect is reversed and the inelastic
scattered intensity decreases as the temperature decreases as
shown in Figs. 1b and 2b. These results are in good agreement with
results reported previously \cite{15}, but that experiment had
coarser resolution and low energy measurements were only possible
for q=0.05~a*.

We emphasize that below T$_{d}$ not only is the intensity of the
low energy part (2$<$E$<$7~meV) decreasing, but the intensity of
the TO1 mode is also diminishing. This effect is especially
pronounced at q=0.05~a* (Fig.~\ref{fig:2}b). The observed
temperature dependent behavior is not attributed to the softening
of the TO1 mode for the following reasons: i) the experimental
curves at q=0.05~a* (Fig.~\ref{fig:2}) and 0.075 a*
(Fig.~\ref{fig:3}) cannot be fit with a single resonance (TO1) or
as a sum of a TA or TO resonance. ii) At q=0.075a* an additional
excitation is visible at E$\approx$4.0~meV. iii) No increase of
inelastic scattering intensity of the TO1 mode at q=0 or the
central peak was reported in Ref.~\onlinecite{17} for (2,0,0)
Brillouin zone.

\begin{figure}
\includegraphics [width=\columnwidth, height= \columnwidth, clip=] {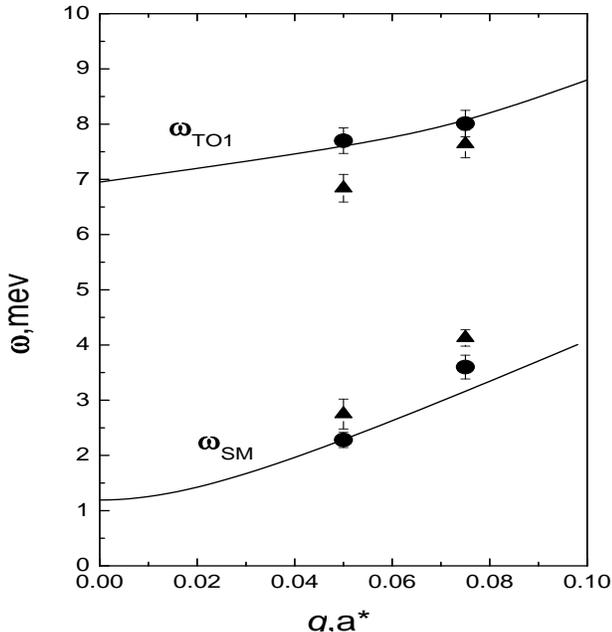}
\caption{Triangles -- frequencies of SM and TO1 modes with mode
coupling neglected; black circles -- obtained from fits to the
data using \protect{Eq.~\ref{eq:i}} at 880 K. lines -- dispersion
curves from \protect{Ref.~\onlinecite{15}}.} \label{fig:4}
\end{figure}

\begin{figure}
\includegraphics [width=\columnwidth, height= 0.9\columnwidth, clip=] {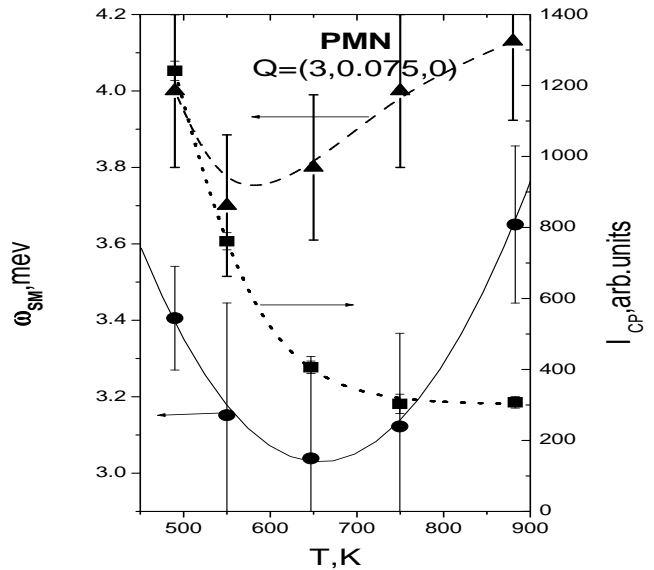}
\caption{Temperature dependence of the SM frequency (triangles --
mode coupling neglected; black circles -- mode coupling included)
and the central peak intensity (squares) at q=0.075a*; lines are
guides for eye.} \label{fig:5}
\end{figure}

The above results clearly indicate the existence in PMN of a
strong temperature dependent intensity associated with a soft
excitation. The results at q=0 (Fig.~\ref{fig:1}) are difficult to
analyze because of the contribution of the Bragg peak intensity at
$\omega $=0, which is several orders of magnitude stronger than
the inelastic part. The data at q=0.05 and 0.075 a* were analyzed
in two different ways. First we fit the data as a sum of 2
gaussians corresponding to the elastic background and the
so-called Bragg tail (result of the long axis of the 4-D
resolution function touching the q=0, w=0 Bragg point) and 2
independent lorentzians, corresponding to the TO1 and soft mode
(the QO mode referred to in Ref.~\onlinecite{15}).  The fit was
successful, but the frequency of the TO1 mode was lower than that
determined in Refs.~\onlinecite{15,17}. Figure~\ref{fig:4} shows
the q dependence of the energy of the modes determined in this way
and Fig.~\ref{fig:5} gives the temperature dependence of the
energy of the SM. Such a discrepancy can be attributed to
neglecting the coupling between the TO1 and SM. In the next stage
we repeated the data analysis using the coupled mode analysis as
done previously \cite{15,20}. The scattered intensity can then be
written as: \noindent
\begin{equation} \label{eq:i}
I(\omega )=[n(\omega) +1]\sum _{ik}(2\omega _i)^{1/2}S_i \,
(2\omega _k)^{1/2}S_k \, {\rm Im} G_{ik} (\omega ),
\end{equation}
\noindent with structure factors $S_{i}$ and the retarded Green's
functions $G_{ik}$ determined by the Dyson equations:

\begin{equation} \label{eq:dyson}
G_{ik}=G_{ik}^0 -\sum_{l,m}G_{il}^0\Pi _{lm}G_{mk} \>,
\end{equation}

\noindent
 The coupling was considered to be linear and classical
harmonic oscillators were used as a zero order approximation. Such
choice is well justified for the TO1 mode. In the case of the soft
mode (SM) it is not clear if the damped harmonic oscillator
represents a proper choice. However, we believe that for our set
of experimental data, utilization of a more complicated expression
(like Lifshitz tail of the density of states used in \cite{21}) is
not adequate here. Using the self energy in the form used in
\cite{15}:

\begin{equation} \label{eq:pi}
\Pi = \left( \begin{array}{cc} -i\frac{\omega}{\omega_1} \gamma_1
&
\Delta_{12}-i\frac{\omega}{\sqrt{\omega_1\omega_2}}\gamma_{12}\\
% \ &\ \\
\Delta_{12}-i\frac{\omega}{\sqrt{\omega_1\omega_2}}\gamma_{12}&
-i\frac{\omega}{\omega_2} \gamma_2
\end{array} \right) \>.
\end{equation}

\noindent where $\omega _{1}$ corresponds to the soft mode (SM)
and $\omega _{2}$ to the TO1 mode are the bare mode frequencies.
$\gamma _{i }$ (i=1,2) are the mode damping constants and $\Delta
_{12}$ and $\gamma _{12}$ are the real and imaginary parts of the
coupling constants.

This model has 8 parameters, 7 of these are independent and 1 is
arbitrary. In the previous study \cite{15} the best agreement for
the description of the coupling between the TA and SM modes has
been achieved under the assumption of purely imaginary coupling,
i.e. $\Delta _{12}$=0. Following Ref.~\onlinecite{22} we treated
the present data under the same conditions. The experimental
spectra were fit by the sum of a quasielastic component with a
gaussian line shape and resolution limited line width, plus a term
described by Eq.~\ref{eq:i} convoluted with the calculated energy
resolution and a Bragg tail . The contribution of the TA phonon
was neglected since it is nearly unobservable in this BZ.

The fit of the data at q=0.05a* was successful at 880 K, but at
lower temperatures the softening resulted in overlapping of the SM
component with the Bragg tail at E$\approx $1.17~meV so below 880K
all parameters were held fixed except for the intensities of the
components. For q=0.075a* the fits were successful at all
temperatures with typical $\chi^{2}$ between 1.5 and 2. The
results are shown in Figs.~\ref{fig:4}
\noindent and \ref{fig:5}. In Fig.~\ref{fig:4} the frequencies of
the TO1 mode and the SM, together with the dispersion curve of the
TO1 mode determined in Ref.~\onlinecite{15}. As can be clearly
seen the present results are in good agreement with the data
obtained from Ref.~\onlinecite{15} determined from the analysis of
the coupling with the TA phonons. In Fig.\ref{fig:5}, the
temperature dependence of the frequency of the SM at q=0.075a* is
shown. Over the temperature range studied the frequency of the SM
varies by $\approx$20{\%} and shows a minimum at $\approx$650~K
which coincides with the Burns temperature determined in this
system \cite{23}. The central peak intensity, also shown in Fig.
5, begins to grow at this same temperature.

\section{Discussion}\label{sec:dis}

\begin{figure}
\includegraphics [width=\columnwidth, height= 0.9\columnwidth, clip=] {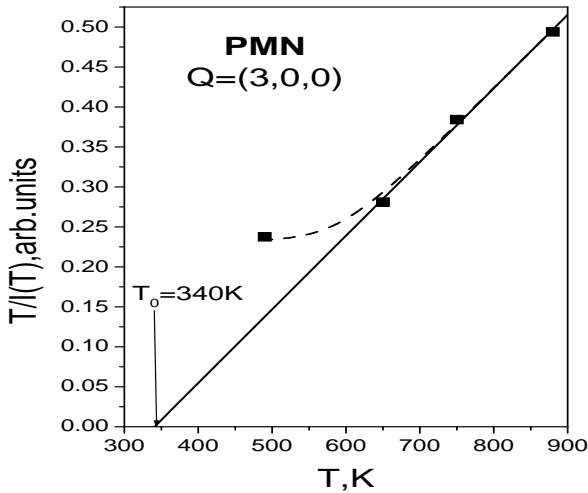}
\caption{T/I(T), where I(T) -- the integrated intensity measured
at Q=(3,0,0) (\protect{Fig. \ref{fig:1}a)} for 2$ < \omega  <
$12~meV vs temperature.} \label{fig:6}
\end{figure}

The results presented above show that for T$>$T$_{d}$ a soft
underdamped excitation exists for q$ \ne $0. This excitation is
clearly resolved for q=0.075~a* in the temperature region studied:
490K$ < $T$ < $880~K. The temperature variation is 20{\%} over
this region and it shows a minimum corresponding to the Burns
temperature T$_{d}\approx$650K. It is important to emphasize that
this mode is distinct from the TO1 mode, which is at a higher
energy. The main argument against the TO1 mode being the soft mode
is the temperature dependence of the dielectric constant,
$\varepsilon $(T). As mentioned above at high temperatures
$\varepsilon $(T) follows Curie-Weiss law with T$_{c}$=398K
\cite{33}. Below the Burns temperature T$_{d}$, the phonon related
part of $\varepsilon $ slowly decreases \cite{18}. Based upon
Lyddane-Sachs-Teller (LST) relationship one expects to see a soft
q=0 mode with energy, $\omega _{0}^{2}$, mirroring the
$\varepsilon $(T) dependence, i. e, $\varepsilon $(T) $\sim $
1/$\omega _{{\rm 0}}^{2}$. We cannot determine the $\omega
_{0}$(q=0) temperature dependence quantitatively because  the mode
is highly damped, Fig.~\ref{fig:1}a. However, the integrated
intensity for phonon scattering in the high temperature limit
($kT>>h\omega )$ varies as I(T) $\sim $ T/$\omega _{0}^{2}$. If we
integrate the intensity for q=0 in Fig.~\ref{fig:1}a (for $\omega
>$2.0~meV to eliminate the elastic part) and plot its reciprocal
(Fig. \ref {fig:6}) \noindent
\noindent we find a linear region above T$_{d}$ that extrapolates
to zero at T$_{0}$ = 340~K. This is in close agreement with the
dielectric measurements\cite{33}, which show a Curie-Weiss
behavior with T$_{0}$=398K. The TO1 mode alone, studied in
Ref.~\onlinecite{17}, cannot account for the dielectric behavior
because it is only weakly temperature dependent and the T$_{0}$
obtained from Ref.~\onlinecite{17} is negative. Thus, the
inelastic response measured about (3,0,0) is the soft mode and can
account for the high temperature dielectric behavior via the LST
relation.

The hardening of the soft mode below T$_{d}$ is also coherent with
the results of ultrabroadband dielectric spectroscopy. Below
T$_{d}$ the dielectric response is determined by the slow polar
cluster dynamics \cite{18} and in neutron inelastic scattering
data it is observed by the appearance of a central peak which
dominates the low-frequency spectra at low temperatures.

The presence of an extra mode is also reasonable in light of the
recent important work of Hirota et al.\cite{34}. They introduced
the concept of a ``phase-shifted condensed mode'' to explain the
observation of strong elastic diffuse scattering below T$_{d}$ in
PMN around the (3,0,0) zone center and its absence near (2,0,0).
Since the diffuse scattering is a result of the condensing of
polar nanoregions, it is reasonable to observe precursor dynamical
behavior associated with this condensation in analogy with soft
mode displacements in normal ordered systems undergoing structural
phase transitions. The extra mode reported here is the phase
shifted soft mode that is responsible for the high temperature
dielectric properties.

The fundamental question remains as to the origin of this extra
mode. We speculate that it is due to the inherent disorder in this
system due to the mixed occupation of B site by Mg and Nb, which
is also the origin of the local polar nanoregions. At small q, in
addition to the normal vibrational modes associated with the
average lattice, new modes can appear due to the mixed crystal
nature, resulting in a situation similar to the two-mode behavior
observed in solid solutions\cite{24,35}. Such a two mode
interpretation of the optic measurements was proposed nearly 20
years ago by Burns and Dacol\cite{23}. Moreover, extra modes were
recently observed in a mixed crystal of a new superconductor
Y$_{0.5}$Lu$_{0.5}$Ni$_{2}$B$_{2}$C \cite{25}. Clearly a more
complete lattice dynamical study of PMN is necessary.

We should note that Nb$^{5 + }$ is a ferroelectric active ion
while Mg$^{2 + }$ is not. Thus one can imagine a soft polar mode
in which only Nb$^{5 + }$ ions participate. Such a collective mode
cannot exist at large q. A first approximation of q$_{max}$, above
which such a Nb mode disappears (becomes localized), can be the
inverse correlation length of chemical order in PMN of about
0.06a* \cite{26}, i.e. of the same order of magnitude as estimated
from the observability of the soft mode \cite{15}. Such an
assumption is in agreement with the results of the present work,
where this mode is observed at q=0.075a*, but not at q=0.1a*.

The presence of a minimum of the frequency and the appearance of a
narrow central peak at T$_{d}$ indicates either a special phase
transition or a change of dynamics from soft-mode to cluster
regime since it is clear that below T$_{d}$ clusters of the polar
phase arise. Different models \cite{27,28,29,30,31,32} of this
phenomenon have already been discussed. Probably the main unsolved
question is topological: are there infinite clusters below the
Burns temperature? In the framework of the present study we can
make no conclusion on this topic.

\section*{Summary}

The main results of the present study are:

For the first time in a relaxor ferroelectric we have directly observed a
soft ferroelectric mode associated with the high temperature Curie-Weiss
law. This mode is clearly distinct from the TO1 mode.

The temperature dependence of the soft mode frequency measured at finite q
passes through a minimum at the Burns temperature, T$_{d}$, at the same time
a narrow central peak develops in the spectra.

Our results confirm the earlier speculation\cite{15} that at
T$_{d}$ a real or local phase transition occurs\cite{34} resulting
in a crossover of the critical dynamics. Above T$_{d}$ the
dynamics is of displacive type, while below it is governed by a
relaxation of the large polar clusters.

\begin{acknowledgments}
It is a pleasure to acknowledge R. Blinc, B. Dorner, T. Egami, P.
Gehring, M. Glinchuk, V. Sakhnenko, G. Shirane, D. Strauch, V.
Stephanovich, P. Timmonin and B. Toperverg for the many useful
discussions and various suggestions. The PMN single crystal was
provided by the Institute of Physics Rostov-on-Don University.
Work at the Ioffe Institute was supported by the RFBR (grant
02-02-16695) and Russian program ``Neutron Researches of Solids''.
Work at BNL was supported by the Division of Materials Research,
US Department of Energy under contract No. DE-AC02-98CH10886. This
work was supported by the Collaboration in Basic Science and
Engineering (COBASE) Program of the National Academy of
Science/National Research Council's office for Central Europe and
Eurasia.
\end{acknowledgments}

\end{document}